# Professional Practice in Higher Education: A Case Study in Faculty Training and Development in Brazil

Edileuza de Freitas Miranda de Mendonça [a] and José Gómez-Galán [b]

**Abstract**: One of the most important debates is currently focusing on specifying the training that university teachers must receive for their professional practice. Improving it in higher education is extremely important not only for the scientific production generated by university but also for the adequacy of the training that future graduates will be offered; professionals facing an increasingly demanding labour market with new needs. University teachers training, thus, should be a priority in academic policies due to their influence and the role played in the evolution of society, as well as being the basis of the quality of Higher Education. This research, a case study in the State of Goias (Brazil), is focused on a sample of practicing university teachers from different fields of knowledge, and has as main objective to know the characteristics of the training received for their professional practice, both in the field of scientific education and their educational role and ability to transfer knowledge. The methodology used has been mixed, not experimental and descriptive, with the help of instruments for data collection and analysis of quantitative and qualitative nature (questionnaires, interviews, monitoring, checklists, documentary analysis, etc.). The results confirmed the initial hypothesis, which stated that university teachers´ current training is primarily scientific and technical, and has gaps in the teacher training required today for a more effective work in the classroom. In this area teachers often use teaching methodologies supported by previous experiences with traditional features. Finally, to optimize this situation, different strategies for university teachers´ training are proposed seeking to improve both their reflection and teaching practice.

[a] Municipal Council of Education of Goiânia, Goiás (Brazil); [b] Universidad Metropolitana, SUAGM (Puerto Rico - United States), Universidad de Extremadura (Spain) Correspondence: Edileuza de Freitas Miranda de Mendonça, Municipal Council of Education of Goiânia, Goiás (Brazil), Rua 104 nº 721 Setor Sul – Goiânia, Goiás (Brazil). efmm@uol.com.br

## 1. Introduction

At the end of the XX Century the Brazilian educational system was reorganized upon the endorsement of the Law of Directives and Bases of National Education Nº 9.394, in use since 1996 (Lei Diretrizes e Bases da Educação Nacional, 1996). After this Law was passed, the statistical data of the educational census of Brazilian higher education, published between 1995-2000 by The National Institute of Pedagogic Studies of Culture and Education Ministry, revealed that specialist academic teachers predominated in the private higher education institutes (INEP, 2001). With effect, the initial education of an academic teacher in Brazil consists of a post-graduate specialization degree called *Latu-sensu* in academic environments. According to the present Law of Directives and Bases of National Education, the *Latu-sensu* degree is a Post-Graduate degree, whereas according to the preceding *Parecer* 977 Law from 1965 it was considered as a Refresher Course (Conselho Nacional de Educação, 1965). Therefore, this is a very important educational process in Brazilian Higher Education. The aim of this current article is to study Brazilian academic teachers' education, with regards to both its scientific and pedagogic scope in order to determine the characteristics of their professional practice and the methodology they use during teaching activities.

Teacher training must be understood as a set of activities that allow teachers to develop abilities and aptitudes in order to improve their performance and action in the academic environment (Nicholls, 2001; Gomez-Galan and Mateos, 2002; Postareff, Lindblom-Ylänne and Nevgi, 2007; Galvão, Reis and Freire, 2011; Ek, Ideland, Jönsson and Malmberg, 2013; Ryan and Ryan, 2013; Gomez-Galan, 2014; Morina, Cortes-Vega and Molina, 2015). Thus, the kind of pedagogic and scientific education that the *latu-sensu* post-graduate program offers to academic teachers during their initial education towards developing their didactic methodology should be questioned. The main purpose and central element of this work is to study the education model achieved in the higher education level processes. From the perspective of a Brazilian academic teacher, they do acquire scientific and technical competence in some of their fields of knowledge during their post graduate education, but according to Cunha (2001), they lack a wider view on how to integrate their knowledge to meet society's needs and also on how to master pedagogic resources which can expose their methodological needs. From the problem studied the assumption is drawn that *latu-sensu* post-graduate courses do not contribute in an effective and adequate way to the scientific and most of all, pedagogic education of the Brazilian academic teacher.

## 2. Methodology

According to what was previously exposed, different goals have been pursued in this study that condition the structure of the theoretical and field research. They are to a) describe the education of academic teachers and their pedagogic organization in the national and international context, from a historical perspective to the present; b) understand the legal organization of the *latu-sensu* post-graduate course as the initial pedagogic educational step of the Brazilian academic teachers; c) demonstrate that economic market interests regulate education processes; and d) acknowledge the influence of traditional and technical education in the attitude of teaching staff with *latu-sensu* post-graduate training. In parallel with these main goals, distinctive secondary goals dependent of the main ones are pursued. They are to verify if academic teachers who have finished their *latu-sensu* post-graduate course consider themselves trained for research and instruction through assessment; to understand the professional profile of the Brazilian academic teacher who graduated under these circumstances; to identify the contributions of the *latu-sensu* post-graduate program in an academic teacher's educational practices; to analyze the curricular directives of the *latu-sensu* post-graduate courses offered in Brazilian higher education institutions; and to suggest changes to the curricular structure of the *latu-sensu* post-graduate course in order to improve the scientific and pedagogic education of Brazilian academic teachers.

The methodology used, therefore, reaches the goals of the study and confirms the validity of the assumption. According to several authors, particularly Teddlie and Tashakkori (2009), Marconi and Lakatos (2010), Cohen, Manion and Morrison (2011) or Bryman (2015), on whose work the field research was based, after an exhaustive bibliographic analysis of the problems it is possible to establish an initial theoretical reference model that helps to determine the changeable elements and elaborate a general plan of study and research. For that reason, the main platform of this research was a scientific literature review, a process of bibliographic and documental research whose main resources were books, scientific articles and legal documents related to the education of academic Brazilian teachers. Of special importance are the results of contributions from Cunha (1989 and 1996), Oliveira (1995), Behrens (1998), Carvalho (2000), Garrido (2002), Demo (2004), Dias da Silva (2005). Gatti (2010), Lima and Donizete (2015), and Silva, Fabro and Duarte (2015). In this manner, regarding the education of the academic teacher, it is considered that the best analysis involving opinions and contributions from others is the analysis of an international environment. Therefore, the work from Wilson, Floden, and Ferrini-Mundy (2001), Samuelowicz and Bain (2001), Beck and Kosnik (2002), Knight (2002), Akerlind (2003), Cochran-Smith and Zeichner (2005), Norton, Richardson, Hartley, Newstead and Mayes (2005), Wilson (2006), Sikes (2006), Postareff, Lindblom-Ylänne and Nevgi (2008), Gunn and Fisk (2013), Gunersel and Etienne (2014), Postareff and Nevgi (2015) or Kauppila (2016) should be highlighted.

In this particular case the work was completed by undertaking theoretical work based on all these scientific contributions related to the study objective, in addition to an analysis of official documents, such as the ranking of Brazilian higher education (INEP, 2001 and 2013) and the curricular directives of 50 *latu-sensu* post-graduate courses in different areas of knowledge (e.g. biology, human sciences, experimental sciences and social sciences) offered by 25 higher education institutions.

The second stage of the research was achieved from the information obtained, which consisted of fieldwork based on the descriptive ex-post-facto method. This involved using standard techniques of opinion study and observation control whose main instruments were questionnaires, lists of controls and interviews. The opinion study sample consisted of 117 academic teachers, 93 of which were from the Catholic University of Goiás State and 24 were from the State University of Goiás State, all of them possessing a *latu-sensu* post-graduate degree. In addition, 40 teachers from the Catholic University of Goiás State who were still undertaking their post-graduate courses were included in the study sample. The observation study included 15 teachers who took part in the State University of Goiás State's University Program for Educational Staff from the Pedagogy graduate course.

### 3. Results

The results obtained from both the theoretical work and from the field work (experimental and theoretical research) can be briefly summarized as follows:

a) In the first stage of the study, according to the results, the education in Brazil is influenced by the North American educative model, independent of all the Humanistic-Christian European influence developed since the Medieval Age and brought to Brazil with the arrival of the Europeans. The first Law of Directives and Principals from 4.024/1961 did not standardize Brazilian Post Graduate programs until 1965, when Post Graduate programs were organized based on the North American model, under *Parecer* Nº 977 (Conselho Nacional de Educação, 1965) aiming to provide the academic teacher with scientific training. The Agreement of Washington influenced the organization of the Law of Directives and Fundamentals Nº 9.394/96 (Lei Diretrizes e Bases da Educação Nacional, 1996), which organizes curricula according to market demands, reducing costs associated with higher education and transferring Education responsibilities from the public to the private sector. The result of this was an increase in the value of the *latu-sensu* courses, which became a post-graduate course instead of a refresher course. The present Law of Directives and Fundamentals consider post-graduation as the preparation for higher education teaching; therefore, specialization is the minimum requirement for this. This explains how the academic education of higher education teachers lays within the same fundamental principle according to what was shown in this study.

b) Experimental research has been able to determine the professional profile of teachers and the complete characteristics of the *latu-sensu* post-graduate program. This research concludes that the education offered is not sufficient to enable the teachers to adapt didactic methodologies to the needs of the student. In addition to the demand of teachers, it was observed that the teaching methods are based on inductive, expositive techniques, with a very limited use of resources (the blackboard and photocopied sheets being the most used techniques). Nevertheless, there is a clear intention to establish pedagogic attitudes and to show interest for the student's learning by motivating interaction, but such processes have only been achieved from experience and not from acquired education. Lastly, the points previously described were validated with the verification of scientific-pedagogic education described in the curricular directives of *latu-sensu* post-graduate courses offered by Brazilian higher education institutions (as stated before, 50 different curricular directives from 25 institutions from Brazil were analyzed).

Some of the main data obtained in this part of the study are presented schematically:

*First Part. Professional Profile of Specialist Academic Teachers*

| *P.I.D. - Identification- Service Years on the Academic Teaching Profession* | | | | |
|---|---|---|---|---|
| *Service Years* | *Amount of Teachers* | *Xi* | *Pi* | *Percentages* |
| 5 years | 80 | 5 | 0,68 | 68% |
| 10 years | 10 | 10 | 0,09 | 9% |
| 15 years | 10 | 15 | 0,09 | 9% |
| 20 years | 6 | 20 | 0,05 | 5% |
| 25 years | 11 | 25 | 0,09 | 9% |
| Total | 117 | | 100 | 100% |

Table 1. *P.I.D. - Professional Profile of Specialist Academic Teachers: Service Years on the Academic Teaching Profession.*

Regarding the *time of service in the academic teaching profession*, 68% of the teachers examined have 5 years of academic teaching experience, 9% have 10 years, 9 % have 15 years, 5 % have 20 years and 9 % have 25 years.

| *P.I.F. - Post Graduation Degree Completed* | | | |
|---|---|---|---|
| *Academic Degree Completed* | *Sum of Teachers* | *Pi* | *Percentages* |
| Specialization | 103 | 0,88% | 88% |
| Masters | 12 | 0,10% | 10% |
| Doctorate | 02 | 0,2% | 2% |
| *Total* | 117 | 1,00% | 100% |

Table 2. *P.I.F.- Post-Graduation Degree Completed.*

*Degrees Completed*: 88% of the examined teachers are specialists, 10% have a master's degree and 2% have a doctorate degree.

| *A.T.C.1 - Scientific-Pedagogic Attitudes of the Teacher in the Classroom: Exposition Methods* | | |
|---|---|---|
| *Method Types* | *Amount of Classes* | *Percentages* |
| Deductive Methods | 61 | 81% |
| Inductive Methods | 14 | 19% |
| *Total* | 75 | 100% |

Table 3. *A.T.C.1 - Scientific-Pedagogic Attitudes of the Teacher in the Classroom: Exposition Methods.*

*About the Exposition Method*: Teachers used the deductive method in 81% of the classes and the inductive method in only 19% of the classes.

| *A.TC.2 - Scientific-Pedagogic Attitudes of the Teacher in the Classroom: Didactic Resources* | | |
|---|---|---|
| *Resource Type* | *Amount of Classes* | *Percentages* |
| Photocopy of texts, blackboard and chalk-pencil | 43 | 58% |
| Slides and overhead projector | 28 | 37% |
| Video Projection | 4 | 5% |
| *Total* | 75 | 100% |

Table 4. *A.T.C.2 - Scientific-Pedagogic Attitudes of the Teacher in the Classroom: Didactic Resources.*

*Regarding didactic resources***:** Teachers used photocopies of texts, blackboard and chalk in 58% of the classes, slides and overhead projectors in 37% of the classes and video projectors in only 5% of the classes.

*A.T.C.3 – Scientific-Pedagogic Attitudes of the Teacher in the Classroom Regarding Contents Outline Planning*

| Contents Arrangement | Amount of Classes | Percentages |
|---|---|---|
| Introduction, body and conclusion | 56 | 75% |
| Introduction and body | 15 | 20% |
| Introduction and conclusion | 4 | 5% |
| Total | 75 | 100% |

Table 5. *A.T.C.3 - Scientific-Pedagogic Attitudes of the Teacher in the Classroom*

*Contents Outline Planning*: Teachers properly organized contents to be offered in 75% of classes, making an introduction, development thereof and concluding with a summary of the gist studied, but in 20% of the classrooms only an introduction and a general development were made without establishing the most important endpoints, and in 4% of them a long introduction was offered as well as general conclusions without having analyzed the contents.

*A.T.C.4 – Scientific-Pedagogic Attitudes of the Teacher in the Classroom. Contextualization of Contents*

| Contextualization of Contents | Amount of Classes | Percentages |
|---|---|---|
| Only subject-related | 11 | 15% |
| Contextualization with other scientific knowledge areas | 42 | 56% |
| Contextualization with other scientific areas, life experiences and social and current issues | 22 | 29% |
| Total | 75 | 100% |

Table 6. *A.T.C.4 - Scientific-Pedagogic Attitudes of the Teacher in the Classroom. Contextualization of Contents*

*Contextualization of Contents*: in 15% of classes teachers focused solely on content related to their subject; however, some connections were established with other scientific areas of knowledge in 56% of them as well as authors' quotes in a broad context; in 29% of them teachers went beyond and contextualization was carried out not only with other scientific areas but with their own life experiences and / or general current issues or social problems, giving more practical sense to their teaching.

| *A.T.C.5 – Scientific-Pedagogic Attitudes of the Teacher in the Classroom. Didactic Techniques Used* | | |
|---|---|---|
| *Didactic Techniques Used* | *Amount of Classes* | *Percentages* |
| Expository Sessions (Lectures) | 73 | 97% |
| Guided Study | 2 | 3% |
| Debate | 0 | 0% |
| *Total* | 75 | 100% |

Table 7. *A.T.C.5 - Scientific-Pedagogic Attitudes of the Teacher in the Classroom. Didactic Techniques Used*

*Didactic Techniques Used*: in 97% of classes teachers employed only expository sessions (lectures), 3% of teachers conducted studies leading to the development of student self-learning; on the other hand, in the selected sample no classroom was found where the debate was used as a teaching technique.

| *A.T.C.6– Scientific-Pedagogic Attitudes of the Teacher in the Classroom. Care over Learning in Class* | | |
|---|---|---|
| *Care over Learning* | *Amount of Classes* | *Percentages* |
| Interest in learning in class | 53 | 71% |
| Disregard for learning in class | 22 | 29% |
| *Total* | 75 | 100% |

Table 8. *A.T.C.6 - Scientific-Pedagogic Attitudes of the Teacher in the Classroom. Care over Learning in Class*

*Care over Learning in Class*: in 71% of classes teachers showed interest in checking how their students´ learning and / or motivation were producing; in 29% of the classrooms they remained indifferent.

| *A.T.C.7 – Scientific-Pedagogic Attitudes of the Teacher in the Classroom. Teachers Readiness during the Lesson* | | |
|---|---|---|
| *Teachers Readiness during the Lesson* | *Amount of Classes* | *Percentages* |
| Remain seated during the lesson | 32 | 43% |
| Remain standing all the time | 37 | 49% |
| Move around the classroom | 6 | 8% |
| *Total* | 75 | 100% |

Table 9. *A.T.C.7 - Scientific-Pedagogic Attitudes of the Teacher in the Classroom. Teachers Readiness during the Lesson*

*Teachers´ Readiness during the Lesson*: in 43% of classes teachers remained seated during the lesson, while 49% were standing but in the same location (next to the desk and / or on stage). Only 8% of teachers observed moved throughout the classroom (and enhanced interaction with students).

*A.T.C.8– Scientific-Pedagogic Attitudes of the Teacher in the Classroom. Time Allocation*

| *Time Allocation* | *Amount of Classes* | *Percentages* |
|---|---|---|
| Remains with the same activity during the lesson | 69 | 92% |
| Deals with several activities during the lesson | 6 | 8% |
| *Total* | 75 | 100% |

Table 10. *A.T.C.8 - Scientific-Pedagogic Attitudes of the Teacher in the Classroom. Time Allocation*

*Time Allocation*: in 92% of classes teachers carried out the same activity throughout the time; only 8% of the observed teachers worked in various activities.

*Second Part. Characteristics of Latu-Sensu Post-Graduation Attended by Academic Teachers*

| *P.I.L. - Description of Fields of Study in the Latu-sensu Post Graduation* | | | |
|---|---|---|---|
| *Fields of Post-Graduation* | *Amount of Teachers* | *Pi* | *Percentages* |
| Human and Social Sciences | 78 | 0,67% | 67% |
| Natural and Applied Sciences | 39 | 0,33% | 33% |
| *Total* | 117 | 1,00% | 100% |

Table 11. *P.I.L. - Characteristics of the Latu-Sensu Post-Graduate Program Attended by Academic Teachers. Description of Fields of Study in the Latu-sensu Post Graduation.*

*Fields of Post-Graduate Study*: 67% attended Human Sciences courses and 33% attended Biological Sciences courses.

| P.I.J. - Duration of Latu-Sensu Post Graduation | | | |
|---|---|---|---|
| *Duration* | *Amount of teachers* | *Pi* | *Percentages* |
| 6 months | 15 | 0,13% | 0,13% |
| 12 months | 60 | 0,51% | 51% |
| 18 months | 18 | 0,15% | 15% |
| 24 months | 24 | 0,21% | 21% |
| *Total* | 117 | 1,00% | 100% |

Table 12. *Duration of Latu-Sensu Post Graduation.*

*Duration of the post-graduation course*: Out of 117 teachers, 51% attended a post-graduate course lasting 12 months, 21% attended a course lasting 24 months, 15% attended a course lasting 18 months and 13% attended a course lasting 6 months.

| P.I.N. – Scientific Monograph Preparation: Requirements and Justification | | | |
|---|---|---|---|
| *Answer* | *Teacher Amount* | *Pi* | *Percentages* |
| Yes | 95 | 0,81% | 81% |
| No | 22 | 0,19% | 19% |
| *Total* | 117 | 1,00% | 100% |

Table 13. *Scientific Monograph Preparation: Requirements and Justification.*

*Requirement of Monograph Preparation and its Justification*: 81% answered "Yes" and 19% answered "No".

In all items, with regard to the inferential analysis, after calculating the Chi square test $\chi^2 = \Sigma\ (\ (f_o - f_e)^2 / f_e\ )$ and contingency coefficient at the junction of nominal and scale variables (gender, age, degree, courses taught, service years, professional status, political affiliation, religion, marital of parental status, national origin), no significant differences in knowledge.

### 4. Discussion and conclusions

The research process confirms the legitimacy of the proposed study hypothesis. Brazilian academic teachers receive basic scientific education in the *latu-sensu* post-graduate program necessary for their teaching practice, but they receive almost no pedagogic education that could be useful during their teaching work. In the same manner, the pursued aims were also achieved. Regarding scientific education, it is believed that technical-scientific abilities are globally acquired from the education received. Compared to the pedagogic education that is highly limited, conditioned by traditional and technical educative concepts, and is based on their own existence.

Based on the results achieved, the present study also obtains other distinctive contributions. For instance, it is established that pedagogic abilities must be developed in the post-graduate course considering that: a) academic teachers must use their scientific abilities in order to teach the students to investigate and not only to transmit information in the classroom; b) other competences such as ethic, moral, social, political and interpretative competences are also important in the education of teachers; c) in order to improve the learning of students dialogue in the classroom should be valued; and d) the post-graduate curricular directives suggest that subjects such as *Methodology of Scientific Research*, *Psychology of the Learning Process* and *Philosophy of Education* (with ethical and communicative content) should be incorporated with the aim of improving the reflection and the pedagogic practice of the teacher.